\documentclass[prl,aps,superscriptaddress,showpacs,nofootinbib]{revtex4}
\usepackage{graphicx,graphics,epsfig}
\usepackage{amsmath}
\usepackage{verbatim}
\usepackage{color}
\usepackage[standard]{ntheorem}

\definecolor{myurlcolor}{rgb}{0,0,0.4}
\definecolor{mycitecolor}{rgb}{0,0.5,0}
\definecolor{myrefcolor}{rgb}{0.5,0,0}
\usepackage{hyperref}
\hypersetup{colorlinks, linkcolor=myrefcolor,
citecolor=mycitecolor, urlcolor=myurlcolor}

\definecolor{nblue}{rgb}{0.2,0.2,0.8}
\definecolor{ngreen}{rgb}{0.2,0.8,0.2}
\definecolor{nred}{rgb}{0.8,0.2,0.2}
\definecolor{nblack}{rgb}{0,0,0}

\date{\today}

\begin{document}

\title{Sharp Contradiction for Local-Hidden-State Model in Quantum Steering}

\author{Jing-Ling Chen\footnote{Correspondence to
J.L.C (chenjl@nankai.edu.cn).}}
 \affiliation{Theoretical Physics Division, Chern Institute of Mathematics, Nankai University,
 Tianjin 300071, People's Republic of China}
 \affiliation{Centre for Quantum Technologies, National University of Singapore,
 3 Science Drive 2, Singapore 117543}

\author{Hong-Yi Su\footnote{Correspondence to
H.Y.S (hysu@mail.nankai.edu.cn).}}
 \affiliation{Theoretical Physics Division, Chern Institute of Mathematics, Nankai University,
 Tianjin 300071, People's Republic of China}
\affiliation{Department of Physics Education, Chonnam National University, Gwangju 500-757, Republic of Korea}

\author{Zhen-Peng Xu}
 \affiliation{Theoretical Physics Division, Chern Institute of Mathematics, Nankai University,
 Tianjin 300071, People's Republic of China}

\author{Arun Kumar Pati\footnote{Correspondence to
A.K.P (akpati@hri.res.in).}}
 \affiliation{Quantum Information and Computation Group,\\
Harish-Chandra Research Institute, Chhatnag Road, Jhunsi, Allahabad 211 019, India}

\date{\today}


\maketitle

\textbf{
In quantum theory, no-go theorems are important as they rule out the existence  of a particular physical model under consideration.
For instance, the Greenberger-Horne-Zeilinger (GHZ) theorem serves as a no-go theorem for the nonexistence of local hidden variable models by presenting a full contradiction for the multipartite GHZ states. However, the elegant GHZ argument for Bell's nonlocality does not go through for bipartite Einstein-Podolsky-Rosen (EPR) state. Recent study on quantum nonlocality has shown that the more precise description of EPR's original scenario is ``steering", i.e., the nonexistence of local hidden state models. Here, we present a simple GHZ-like contradiction for any bipartite pure entangled state, thus proving a no-go theorem for the nonexistence of local hidden state models in the EPR paradox. This also indicates that the very simple steering paradox presented here is indeed the closest form to the original spirit of the EPR paradox.
}

\vspace{5mm}


In 1935, Einstein, Podolsky and Rosen (EPR) questioned the completeness of quantum mechanics under the
assumption of locality and  reality~\cite{EPR} that underlie the classical world view. By considering continuous-variable entangled state, EPR proposed a famous thought experiment that involves a dilemma concerning local realism against quantum mechanics. This dilemma is nowadays  well-known as the EPR paradox. For a long time, the EPR argument remained a philosophical problem at the foundation of quantum mechanics. In 1964, Bell made an important step forward~\cite{Bell} by considering a version based on the entanglement of spin-1/2 particles introduced by Bohm. The EPR paradox, according to Bell's reasoning, could, supposedly, be resolved by supplementing the theory with local hidden variables (LHV), which nevertheless show an incompatibility with quantum predictions via violation of Bell's inequality. Later, the violation of the so-called Clause-Horne-Shimony-Holt (CHSH) inequality, was verified experimentally~\cite{chsh}.

As for the violation of Bell's inequality, the incompatibility between the LHV models and quantum mechanics was essentially demonstrated in a statistical manner. If instead one aims to achieve a more sharper conflict, one can have the Greenberger-Horne-Zeilinger (GHZ) theorem, an ``all-versus-nothing" proof of Bell's nonlocality that applies to three or more parties~\cite{GHZ89,GHZ90}. The elegant GHZ argument involved the three-qubit GHZ state~\cite{GHZ90}
\begin{eqnarray}
|{\rm GHZ}\rangle=\frac{1}{\sqrt{2}}(|000\rangle+|111\rangle),\label{GHZ}
\end{eqnarray}
where $|0\rangle$ and $|1\rangle$ are the eigenstates of the
Pauli matrix $\sigma_z$ with the eigenvalues $+1$ and $-1$.
respectively. It is easy to verify that the GHZ state is the common eigenstate of the following four mutually commutative
operators: $\sigma_{1x}\sigma_{2x}\sigma_{3x}$, $\sigma_{1x}\sigma_{2y}\sigma_{3y}$, $\sigma_{1y}\sigma_{2x}\sigma_{3y}$, and $\sigma_{1y}\sigma_{2y}\sigma_{3x}$ (here $\sigma_{1x}$\ denotes the Pauli matrix $\sigma_{x}$ measured on the $1$st qubit, similarly for the others),  with the eigenvalues being $+1$, $-1$, $-1$, $-1$, respectively. However, a contradiction arises if one tries to interpret the quantum result with LHV models. Specifically, we denote the supposedly definite values of $\sigma_{1x}$, $\sigma_{2y}$, $...$\ as $v_{1x}$,
$v_{2y}$, $...$ (with $v$'s being 1 or $-1$), then a product of the last three operators, according to LHV models, yields $v_{1y}^{2}v_{2y}^{2}v_{3y}^{2}v_{1x}v_{2x}v_{3x}=-1$, in sharp contradiction to the first operator $v_{1x}v_{2x}v_{3x}=+1$. Such a full contradiction ``$1=-1$" indicates that the GHZ theorem is a \emph{no-go} theorem for quantum nonlocality, i.e., there is no room for the LHV model to completely describe quantum predictions of the GHZ state. The GHZ theorem has already been verified by photon-based experiment~\cite{pan}, and recently a fault-tolerant test of the GHZ theorem  has also been proposed based on nonabelian anyons~\cite{deng}.

In the original formulation of the EPR paradox~\cite{EPR}, a bipartite entangled state is considered which
is a common
eigenstate of the relative position $\hat{x}_1-\hat{x}_2$ and the
total linear momentum $\hat{p}_1+\hat{p}_2$ and can be expressed as
\begin{eqnarray}
\Psi(x_1,x_2)=\int_{-\infty}^\infty e^{(2\pi i/\hbar)(x_1-x_2+x_0)p}
dp, \label{eq1}
\end{eqnarray}
with $\hbar$ the Planck constant. Experimentally one can generate the two-mode squeezed vacuum state
in the nondegenerate optical parametric amplifier (NOPA) \cite{NOPA}
as
\begin{equation}
| {\rm NOPA}\rangle =e^{r(a_1^{\dagger }a_2^{\dagger
}-a_1a_2)}|00\rangle =\sum_{m=0}^\infty \frac{(\tanh r)^m}{\cosh r}%
|mm\rangle ,  \label{nopa}
\end{equation}
where $r>0$ is the squeezing parameter, $\left| mm\right\rangle
\equiv \left| m\right\rangle _1\otimes \left| m\right\rangle _2=\frac 1{m!}%
(a_1^{\dagger })^m(a_2^{\dagger })^m\left| 00\right\rangle $,
$a_i,a_i^{\dagger } (i=1,2)$ are respectively the annihilation and
creation operators, $|m\rangle\equiv|\Psi_m(x)\rangle$ are the Fock
states of the Harmonic oscillator. In the infinite squeezing limit,
$\left| {\rm NOPA}\right\rangle \left| _{r\rightarrow \infty
}\right.=\Psi(x_1,x_2) $, thus the original EPR state is a maximally
entangled state for the bipartite continuous-variable system.


Since the discovery of the EPR paradox,
the question of whether the original EPR state possesses the LHV models has pushed many researchers to achieve intriguing and thought provoking results~\cite{bell,EPR-1,EPR-2,EPR-3,EPR-4,EPR-5}. Bell first showed that the Wigner function of the EPR state, due to its positive definiteness, can directly be used to construct the LHV models \cite{bell}. However, attempt has also been made to reveal its nonlocality in phase space by considering displaced parity operators upon the NOPA state in the large $r$ limit~\cite{EPR-1}. Moreover, maximal violations of the EPR state by multicomponent Bell's inequalities have also been investigated in Refs. ~\cite{chen,chen2}.

Very recently the notion of ``steering"~\cite{WJD07,WJD07PRA} has stimulated people to reconsider the exact implication of the EPR argument. For instance, Werner has remarked on  why Einstein did not go all the way to discover Bell's inequality~\cite{werner}
Steering is indeed  a quite old concept. In response to the EPR paper~\cite{Schrodinger35}, Schr\"odinger, who believed the validity of quantum mechanical descriptions of Nature, introduced in the same year of EPR's paper a term ``steering" to depict the ``spooky action at a distance" which was mentioned in the EPR paper. Specifically, steering in a bipartite scenario describes an ability of one party, say Alice, to prepare the other party's (say Bob's) particle with different quantum states by simply measuring her own particle with different settings. This is also at the heart of remote state preparation protocol using EPR state \cite{akp}. However, 
steering lacked operational meanings, until in the year 2007 Wiseman \emph{et al.}~\cite{WJD07,WJD07PRA} gave a rigorous definition of it through the quantum information task. It then turns out that the EPR paradox concerns more precisely the existence of local hidden state (LHS) models, rather than that of LHV models leading to Bell's inequality.

That is, the \emph{exact} type of quantum nonlocality in the
EPR paradox is EPR steering, rather than Bell nonlocality. After
that, there has been rapid development in EPR steering both
theoretically and experimentally~\cite{SU,QKD,AVN,CS,AVN-EXP}, such as in the
test of steering inequalities~\cite{NP2010,NC,PRX,NJP} and the
experimental observation of one-way EPR steering~\cite{AS12}.

Thus, a natural question arises: since there exist a simple GHZ paradox, i.e.,  ``$1=-1$", which rule out the LHV models more uncompromisingly than Bell inequalities, one may ask whether a similar contradiction can be found so as to completely rule out the LHS models, especially for the EPR state. The merits of confirmatively answering this question include not only finding out the aforementioned missing piece of proofs of steering in analogy to proofs of Bell nonlocality, but also accomplishing the demonstration of \emph{the} EPR paradox in its most original sense.


The aim of this paper is to present a very simple steering paradox, i.e., ``$2=1$", which intuitively demonstrates the steerability for the EPR state, directly confirming that EPR steering is exactly the type of quantum nonlocality inherited in the EPR paradox, henceforth proving a no-go theorem for nonexistence of LHS models in EPR's original sense.


\vspace{5mm}
\noindent{\bf Results}

\noindent{\bf Simple steering paradox in two qubits.} We shall show that in the original EPR's scenario, there exists a simple steering paradox that leads to `$2=1$". A two-setting EPR steering scenario together with a bipartite entangled state are sufficient to demonstrate this full contradiction.

To illustrate the central idea, let us first consider the two-qubit case. In a two-setting steering
protocol of $\{\hat{n}_1, \hat{n}_2\}$ (with $\hat{n}_1\neq
\hat{n}_2$), Alice prepares a two-qubit state $\rho_{AB}$, she keeps
one and sends the other to Bob. Bob asks Alice to perform his choice of either one of two
 possible projective measurements
 (i.e. two-setting) $\mathcal
{P}^{\hat{n}_1}_a$ and $\mathcal {P}^{\hat{n}_2}_a$ on her qubit and
tell him the measurement results of $a$. Here
\begin{equation}
 \mathcal {P}^{\hat{n}}_a=\frac{\openone+(-1)^a {\hat{n}}\cdot {\vec \sigma}}{2}
\end{equation}
is the projector,
with $\hat{n}=(n_x, n_y, n_z)$ the measurement direction, $a$ (with
$a=0,1$) the Alice's measurement result, $\openone$ the $2 \times 2$
identity matrix, and ${\vec \sigma}=(\sigma_x, \sigma_y, \sigma_z)$
the vector of the Pauli matrices. After Alice's measurements, Bob
obtains four conditional states as $\tilde{\rho}^{\hat{n}_j}_a={\rm
tr}_A[(\mathcal {P}^{\hat{n}_j}_a\otimes \openone) \rho_{AB}]$ with
$j=1,2$ and $a=0, 1$. Suppose Bob's state has a LHS description, then there exists an ensemble $\{ \wp_{\xi}
\rho_{\xi} \}$ and a stochastic map $\wp(a|\hat{n},\xi)$ satisfying
\begin{eqnarray}
&&\tilde{\rho}^{\hat{n}}_a=\sum_{\xi} \wp(a|\hat{n},\xi) \wp_{\xi}
\rho_{\xi},\\
&&\sum_\xi \wp_{\xi} \rho_{\xi}=\rho_B, \label{const1}
\end{eqnarray}
where
$\wp_{\xi}$ (with $\wp_{\xi}>0$) and $\wp(a|\hat{n},\xi)$ are probabilities satisfying
$\sum_\xi\wp_{\xi}=1$, and $\sum_a \wp(a|\hat{n},\xi) =1$ for a fixed $\xi$, and
$\rho_B={\rm tr}_A(\rho_{AB}) $ is Bob's reduced density matrix (or Bob's unconditioned state)~\cite{WJD07,WJD07PRA}.

Then, Bob will check the following set of
four equations:
\begin{eqnarray}\label{E0}
 \tilde{\rho}^{\hat{n}_1}_0&=& \sum_\xi \wp(0|\hat{n}_1,\xi) \wp_{\xi} \rho_{\xi},\label{Erhoz0q}\nonumber\\
 \tilde{\rho}^{\hat{n}_1}_1&=&\sum_\xi \wp(1|\hat{n}_1,\xi) \wp_{\xi} \rho_{\xi},\label{Erhoz1q}\nonumber\\
 \tilde{\rho}^{\hat{n}_2}_0&=&\sum_\xi \wp(0|\hat{n}_2,\xi) \wp_{\xi} \rho_{\xi},\label{Erhox0q}\nonumber\\
 \tilde{\rho}^{\hat{n}_2}_1&=&\sum_\xi \wp(1|\hat{n}_2,\xi) \wp_{\xi} \rho_{\xi}.\label{Erhox1q}
\end{eqnarray}
If these four equations have a contradiction (or say they cannot
have a common solution of $\{ \wp_{\xi} \rho_{\xi} \}$ and
$\wp(a|\hat{n},\xi)$), then Bob is convinced that a LHS model does
not exist and Alice can steer the state of his qubit.

Now, let the state $\rho_{AB}$ be an arbitrary two-qubit pure entangled state, which is given in its Schmidt form as
\begin{eqnarray}
 |\Psi(\theta)\rangle=\cos\theta|00\rangle+\sin\theta|11\rangle,
 \end{eqnarray}
where $\theta\in (0, \pi/2)$.
The pure entangled state
$\rho_{AB}=|\Psi(\theta)\rangle\langle\Psi(\theta)|$ has a
remarkable property: Bob's normalized conditional states
${\rho}^{\hat{n}}_a=\tilde{\rho}^{\hat{n}}_a/{\rm
tr}(\tilde{\rho}^{\hat{n}}_a)$ are always pure, and $\{{\rho}^{\hat{n}_1}_0, {\rho}^{\hat{n}_1}_1\}\neq \{{\rho}^{\hat{n}_2}_0, {\rho}^{\hat{n}_2}_1\}$ for $\hat{n}_1\neq
\hat{n}_2$ (Here ${\rho}^{\hat{n}_1}_0, {\rho}^{\hat{n}_1}_1, {\rho}^{\hat{n}_2}_0, {\rho}^{\hat{n}_2}_1$ are four different pure states when $\rho_{AB}$ is a pure entangled state).
It is well-known
that a pure state cannot be obtained by a convex sum of other
different states, namely, a density matrix of pure state can only be
expanded by itself. Therefore without loss of generality, from
Eq. (\ref{E0}) one has
\begin{eqnarray}\label{four}
\tilde{\rho}^{\hat{n}_1}_0=\wp_{1} \rho_{1},\nonumber\\
\tilde{\rho}^{\hat{n}_1}_1=\wp_{2} \rho_{2},\nonumber\\
\tilde{\rho}^{\hat{n}_2}_0=\wp_{3} \rho_{3},\nonumber\\
\tilde{\rho}^{\hat{n}_2}_1=\wp_{4} \rho_{4},
\end{eqnarray}
with the probabilities $\wp(0|\hat{n}_1,\xi=1)=\wp(1|\hat{n}_1,\xi=2)=\wp(0|\hat{n}_2,\xi=3)=\wp(1|\hat{n}_2,\xi=4)=1$, and other terms are zeros (see \textbf{Methods} for more detail of derivation).
By summing them up and taking trace, due to $\rho_B={\rm
tr}_A(\rho_{AB})=\tilde{\rho}^{\hat{n}}_0+\tilde{\rho}^{\hat{n}}_1$,
the left-hand side gives $2 {\rm tr}\rho_B=2$. But the right-hand
side, by definition, gives ${\rm tr}(\sum_{\xi=1}^4 \wp_{\xi} \rho_{\xi})={\rm tr}\rho_B=\sum_{\xi=1}^4 \wp_{\xi}=1$, this leads to
a full contradiction of ``$2=1$".


The above simple paradox ``$2=1$" offers a transparent argument
of nonexistence of LHS models (or existence of EPR steering) for a two-qubit pure entangled state. The subtlety
of the paradox lies in the fact the wavefunction
$|\Psi(\theta)\rangle$ can have different decompositions, such as
\begin{eqnarray}
 |\Psi(\theta)\rangle&=&\cos\theta|0\rangle|0\rangle+\sin\theta|1\rangle|1\rangle\nonumber\\
 &=&\frac{1}{\sqrt{2}}( |+\rangle|\chi_+\rangle+|-\rangle|\chi_-\rangle ),
 \end{eqnarray}
with $|\pm\rangle=\frac{1}{\sqrt{2}}(|0\rangle\pm|1\rangle)$ and
$|\chi_{\pm}\rangle=\cos\theta|0\rangle\pm\sin\theta|1\rangle$.
In practice, the two-setting protocol can be chosen as $\{\hat{z}, \hat{x}\}$. Namely, Bob asks Alice to measure
her qubit along the $\hat{z}$-direction and  the $\hat{x}$-direction, respectively.
Suppose Alice performs her measurement in the $\hat{z}$-direction (or the $\hat{x}$-direction), for convenient,
one may denote the set of her projectors as $\mathbb{Z}=\{
|0\rangle\langle 0|,|1\rangle\langle 1|\}$ (or $\mathbb{X}=\{
|+\rangle\langle +|,|-\rangle\langle -|\}$), then she can
project Bob's system into one of the pure states
$\{|0\rangle,|1\rangle\}$ (or $\{|\chi_+\rangle,|\chi_-\rangle\})$.
It is easy to verify that $\mathbb{Z},\mathbb{X}$ are locally orthogonal and complete
bases. Namely, $\langle0|1\rangle=\langle+|-\rangle=0$, $\sum_{m=0,1}|m\rangle\langle m|=\sum_{m'=+,-}|m'\rangle\langle m'|=\openone$, and the basis $\mathbb{X}$ can be obtained from the diagonal basis $\mathbb{Z}$ through a unitary transformation.

\vspace{5mm}
\noindent{\bf Generalization to bipartite high-dimensional systems.} Suppose in the steering scenario, the quantum state that Alice prepares is a pure entangled state of two $d$-dimensional systems (two-qudit), then one can have the same simple paradox ``$2=1$".

Let us consider the two-qudit pure entangled state in its Schmidt form
\begin{eqnarray}
 |\Phi\rangle=\sum_{m=0}^{d-1} \lambda_m |mm\rangle,
 \end{eqnarray}
where $|m \rangle$ is the state in the diagonal basis, $\lambda_m$'s are the Schmidt coefficients, and $\sum_{m=0}^{d-1} \lambda_m^2=1$. In the two-setting steering
protocol of $\{\mathbb{Z},\mathbb{X}\}$, Alice prepares a two-qudit pure state $\rho_{AB}= |\Phi\rangle\langle  \Phi|$, she keeps one and sends the other to Bob. To verify the steerablity of Alice, Bob asks Alice to perform his choice of either one of two
 possible projective measurements $|m \rangle\langle m|$ and $|m' \rangle\langle m'|$ on her qubit and
tell him the measurement results of $m$ and $m'$. Similarly, the sets of projectors for Alice are as follows
\begin{eqnarray}
 \mathbb{Z}&=&\{|m \rangle\langle m| | m=0,1,2,...,d-1\},\nonumber\\
 \mathbb{X}&=&\{|m' \rangle\langle m'| | m'=0,1,2,...,d-1\}.
 \end{eqnarray}
In principle, the choice of $\mathbb{Z}$ and $\mathbb{X}$ is rather arbitrary, as long as any element in $\mathbb{Z}$ does not fully overlap with that in $\mathbb{X}$. For simplicity $\mathbb{Z}$ and $\mathbb{X}$ here can be taken as two of the mutually unbiased bases for a $d$-dimensional system, such that $|\langle m|m'\rangle|^2=1/d$ for any pair of $m$ and $m'$.
After Alice's measurements, Bob
obtains $2d$ conditional states as $\tilde{\rho}^{\mathbb{Z}}_m={\rm
tr}_A[(|m \rangle\langle m|\otimes \openone) \rho_{AB}]$ and $\tilde{\rho}^{\mathbb{X}}_{m'}={\rm
tr}_A[(|m' \rangle\langle m'|\otimes \openone) \rho_{AB}]$. Similarly, Bob can check the following
 set of $2d$ equations:
\begin{eqnarray} \label{newE0}
 \tilde{\rho}^{\mathbb{Z}}_m&=& \sum_\xi \wp(m|\mathbb{Z},\xi) \wp_{\xi} \rho_{\xi},\label{newErhoz0q}\nonumber\\
 \tilde{\rho}^{\mathbb{X}}_{m'}&=& \sum_\xi \wp(m'|\mathbb{X},\xi) \wp_{\xi} \rho_{\xi},\label{newErhox0q}
\end{eqnarray}
with $m, m'=0,1,2,...,d-1$.
If these $2d$ equations have a contradiction, then there is no a LHS model description and Bob has to be convinced that Alice can steer the state of his qubit.

Because $\rho_{AB}=|\Phi\rangle\langle\Phi|$ is a pure entangled state, it can be directly verified that
Bob's normalized conditional states
are always pure, for instance one has ${\rho}^{\mathbb{Z}}_m=\tilde{\rho}^{\mathbb{Z}}_m/{\rm tr} \tilde{\rho}^{\mathbb{Z}}_m=|m \rangle\langle m|$.
Due to the fact that a density matrix of pure state can only be
expanded by itself, therefore, from
equation (\ref{newE0}) one has
\begin{eqnarray}
\tilde{\rho}^{\mathbb{Z}}_m&=&\wp_{m} \rho_{m},\nonumber\\
\tilde{\rho}^{\mathbb{X}}_{m'}&=&\wp_{m'} \rho_{m'},
\end{eqnarray}
with $m, m'=0,1,2,...,d-1$.
By summing them up and taking the trace, we have
\begin{eqnarray}
\rho_B&=&{\rm tr}_A(\rho_{AB})\nonumber\\
&=&\sum_{m=0}^{d-1} \tilde{\rho}^{\mathbb{Z}}_m=\sum_{m'=0}^{d-1} \tilde{\rho}^{\mathbb{X}}_{m'}\nonumber\\
&=&\sum_{m=0}^{d-1} \wp_{m} \rho_{m}+\sum_{m'=0}^{d-1} \wp_{m'} \rho_{m'}.
\end{eqnarray}
From (15), one sees that the left-hand side gives $2 {\rm tr}\rho_B=2$ and the right-hand
side gives ${\rm tr}\rho_B=1$,  leading to a full contradiction of ``$2=1$".

The above analysis is also valid when $d$ tends to infinity. By chosing $\lambda_m= \frac{(\tanh r)^m}{\cosh r}$ and let $d\rightarrow \infty$, then one can have a similar paradox ``$2=1$" for the continuous-variable state $| {\rm NOPA}\rangle$, which includes the original EPR state by taking the infinite squeezing limit. Thus, we complete the demonstration of the simple steering paradox for the original EPR scenario, which is a no-go theorem for nonexistence of LHS models in the EPR paradox. In other words, the sharp contradiction ``$2=1$" indicates that there is no room for the LHS description of any bipartite pure entangled state, including the original EPR state.

\emph{Remark 1.}---The original EPR state has the following elegant decompositions
\begin{eqnarray}
|\Psi\rangle_{\rm
EPR}&=&\sum_{m=0}^{\infty}c_m|\Psi_m(x_1)\rangle|\Psi_m(x_2)\rangle\nonumber\\
&=&\sum_{m=0}^{\infty}c_m|\Psi_m(x_1+\ell)\rangle|\Psi_m(x_2+\ell)\rangle,
\end{eqnarray}
where in the last step we have operated a translation transformation
$e^{i(p_1+p_2)\ell/\hbar}$ on $|\Psi\rangle_{\rm EPR}$ that does not change the state $|\Psi\rangle_{\rm EPR}$,  $\ell$ is a real number, and
$|\Psi_m(x_i+\ell)\rangle=e^{ip_i\ell/\hbar} |\Psi_m(x_1)\rangle$. Thus the two-setting steering protocol can be chosen as $\{ |\Psi_m(x_1)\rangle\langle \Psi_m(x_1)|, |\Psi_m(x_1+\ell)\rangle\langle \Psi_m(x_1+\ell)| \}$.

\vspace{5mm}
\noindent\textbf{Discussions}

\noindent
The EPR paradox has resulted in search for local hidden variable models with locality and reality as starting points, but Bell's inequaliy rules out such mdels as the predictions of LHV models do not match quantum theory.
The GHZ paradox demonstrates sharp contradiction between the predictions of local hidden variable theory and quantum mechanics without using any inequality. However,  the  GHZ paradox is not applicable to bipartite systems. Hardy did attempt to extend the all-versus-nothing argument to a two-qubit system to reveal Bell's nonlocality~\cite{hardy,hardy-2}, and this proof is usually considered as ``the best version of Bell's theorem"~\cite{mermin}.
However, Hardy's proof works for only $9\%$ of the runs of a specially constructed experiment, and moreover, it is not
valid for two-qubit maximally entangled state. Thus, in this sense, Hardy's proof may not  be considered appropriately as the closest form to the spirit of EPR's original scenario.

In summary,  we have presented a simple steering paradox  that shows the incompatibility of the local hidden state model with quantum theory for any bipartite pure entangled state, including the original EPR state. The full contradiction that results in $``2=1"$ not only  intuitively demonstrates the steerability for the EPR state, directly confirming that EPR steering is exactly the type of quantum nonlocality inherited in the EPR paradox, but also indicates that the very simple steering paradox is the closest in its form to the spirit of the EPR paradox. Furthermore, if one considers the EPR steering scenario in $k$-setting, then following the similar derivation one can arrive at a full contradiction , i.e., $``k=1"$. We expect that the simple steering paradox can be demonstrated in both two-qubit system and continuous-variable system by photon entangled based experiments in the near future.



\vspace{5mm}
\noindent \textbf{Methods}
\vspace{5mm}

\noindent{\bf Detail derivation of the steering paradox for  two qubits.} It can be directly verified that, if the state $\rho_{AB} = |\Psi(\theta)\rangle\langle\Psi(\theta)|$ is a pure entangled state, then  ${\rho}^{\hat{n}_1}_0, {\rho}^{\hat{n}_1}_1, {\rho}^{\hat{n}_2}_0, {\rho}^{\hat{n}_2}_1$ are four different pure states. For example and for convenient, let us take
\begin{eqnarray}
\hat{n}_1=\hat{z}, \;\;\;\;\; \hat{n}_2=\hat{x}.
\end{eqnarray}
Then in the two-setting steering protocol of $\{\hat{z}, \hat{x}\}$, Bob asks Alice to perform his choice of either one of two
 possible projective measurements along the $z$-direction (with the projector  $\mathcal
{P}^{\hat{z}}_a$) and the $x$-direction (with the projector $\mathcal
{P}^{\hat{x}}_a$) on her qubit and
tell him the measurement results of $a$ (with $a=0, 1$). More precisely, one has the projectors as
\begin{eqnarray}
 \mathcal {P}^{\hat{z}}_0&=& |0\rangle\langle 0|, \nonumber\\
 \mathcal {P}^{\hat{z}}_1&=& |1\rangle\langle 1|, \nonumber\\
 \mathcal {P}^{\hat{x}}_0&=& |+\rangle\langle +|, \nonumber\\
 \mathcal {P}^{\hat{x}}_1&=& |-\rangle\langle -|,
\end{eqnarray}
with $ |\pm\rangle=\frac{1}{\sqrt{2}}(|0\rangle \pm|1\rangle)$.
Then Bob's four unnormalized conditional states become
\begin{eqnarray}
 \tilde{\rho}^{\hat{z}}_0&=& {\rm
tr}_A[(\mathcal {P}^{z}_0\otimes \openone) \rho_{AB}]=\cos^2\theta |0\rangle\langle 0|,\label{Erhoz0q-1}\nonumber\\
 \tilde{\rho}^{\hat{z}}_1&=&{\rm
tr}_A[(\mathcal {P}^{z}_1\otimes \openone) \rho_{AB}]=\sin^2\theta |1\rangle\langle 1|,\label{Erhoz1q-1}\nonumber\\
 \tilde{\rho}^{\hat{x}}_0&=&{\rm
tr}_A[(\mathcal {P}^{x}_0\otimes \openone) \rho_{AB}]=\frac{1}{2} |\chi_+\rangle\langle \chi_+|,\label{Erhox0q-1}\nonumber\\
 \tilde{\rho}^{\hat{x}}_1&=&{\rm
tr}_A[(\mathcal {P}^{x}_1\otimes \openone) \rho_{AB}]=\frac{1}{2} |\chi_-\rangle\langle \chi_-|,\label{Erhox1q-1}
\end{eqnarray}
with $|\chi_\pm\rangle=\cos\theta|0\rangle\pm\sin\theta|1\rangle$.
Thus, Bob's four normalized conditional states are
\begin{eqnarray}
 {\rho}^{\hat{z}}_0&=&  |0\rangle\langle 0|,\nonumber\\
 {\rho}^{\hat{z}}_1&=& |1\rangle\langle 1|,\nonumber\\
 {\rho}^{\hat{x}}_0&=& |\chi_+\rangle\langle \chi_+|,\nonumber\\
 {\rho}^{\hat{x}}_1&=& |\chi_-\rangle\langle \chi_-|,
\end{eqnarray}
which are obviously four different pure states.

Now, if Bob's four unnormalized conditional states can have a LHS description, then they must satisfy
\begin{eqnarray}
 \tilde{\rho}^{\hat{z}}_0&=&\sum_\xi \wp(0|\hat{z},\xi) \wp_{\xi} \rho_{\xi},\label{Erhoz0q-2}\\
 \tilde{\rho}^{\hat{z}}_1&=&\sum_\xi \wp(1|\hat{z},\xi) \wp_{\xi} \rho_{\xi},\label{Erhoz1q-2}\\
 \tilde{\rho}^{\hat{x}}_0&=&\sum_\xi \wp(0|\hat{x},\xi) \wp_{\xi} \rho_{\xi},\label{Erhox0q-2}\\
 \tilde{\rho}^{\hat{x}}_1&=&\sum_\xi \wp(1|\hat{x},\xi) \wp_{\xi} \rho_{\xi}.\label{Erhox1q-2}
\end{eqnarray}
Since the four states in the left-hand-side of Eqs. (\ref{Erhoz0q-2})-(\ref{Erhox1q-2}) are all proportional to pure states, thus it is sufficient for $\xi$ to run from 1 to 4, namely, one can take the ensemble as
\begin{eqnarray}
\{ \wp_{\xi} \rho_{\xi} \}=\{ \wp_1 \rho_1, \wp_2 \rho_2, \wp_3 \rho_3, \wp_4 \rho_4 \},
\end{eqnarray}
with $\wp_i>0$ (if $\wp_{\xi'}=0$, it implies that the corresponding state $\rho_{\xi'}$ is not the hidden state
 considered in the ensemble $\{ \wp_{\xi}
\rho_{\xi} \}$), and $\rho_i$ ($i=1, 2, 3, 4$) are the hidden states.
Then,  Eqs. (\ref{Erhoz0q-2})-(\ref{Erhox1q-2}) become
\begin{widetext}
\begin{eqnarray}\label{E0-p3}
 \tilde{\rho}^{\hat{z}}_0&=& \wp(0|\hat{z},1) \wp_1 \rho_1+\wp(0|\hat{z},2) \wp_2 \rho_2+\wp(0|\hat{z},3) \wp_3 \rho_3+\wp(0|\hat{z},4) \wp_4 \rho_4,\label{Erhoz0q-3}\\
 \tilde{\rho}^{\hat{z}}_1&=&\wp(1|\hat{z},1) \wp_1 \rho_1+\wp(1|\hat{z},2) \wp_2 \rho_2+\wp(1|\hat{z},3) \wp_3 \rho_3+\wp(1|\hat{z},4) \wp_4 \rho_4,\label{Erhoz1q-3}\\
 \tilde{\rho}^{\hat{x}}_0&=&\wp(0|\hat{x},1) \wp_1 \rho_1+\wp(0|\hat{x},2) \wp_2 \rho_2+\wp(0|\hat{x},3) \wp_3 \rho_3+\wp(0|\hat{x},4) \wp_4 \rho_4,\label{Erhox0q-3}\\
 \tilde{\rho}^{\hat{x}}_1&=&\wp(1|\hat{x},1) \wp_1 \rho_1+\wp(1|\hat{x},2) \wp_2 \rho_2+\wp(1|\hat{x},3) \wp_3 \rho_3+\wp(1|\hat{x},4) \wp_4 \rho_4.\label{Erhox1q-3}
\end{eqnarray}
\end{widetext}
In the following, we come to show a simple steering paradox ``$2=1$" based on Eqs. (\ref{Erhoz0q-3})-(\ref{Erhox1q-3}) under the constraints of Eqs. (\ref{const1}), and
\begin{eqnarray}\label{const2}
\wp_{\xi}>0, \;\;\;\;\;
\sum_\xi\wp_{\xi}=1,
\end{eqnarray}
\begin{eqnarray}\label{const3}
\sum_a \wp(a|\hat{n},\xi) =1.
\end{eqnarray}

It is well-known
that a pure state cannot be obtained by a convex sum of other
different states, namely, a density matrix of pure state can only be
expanded by itself. Let us look at Eq. (\ref{Erhoz0q-3}), because the left-hand side is proportional to a pure state, without loss of generality, one has
\begin{eqnarray}
&& \tilde{\rho}^{\hat{z}}_0= \wp(0|\hat{z},1) \wp_1 \rho_1,\nonumber\\
 && \wp(0|\hat{z},2)=\wp(0|\hat{z},3)=\wp(0|\hat{z},4)=0.
 \end{eqnarray}
Similarly, one has
\begin{eqnarray}
&&  \tilde{\rho}^{\hat{z}}_1=\wp(1|\hat{z},2) \wp_2 \rho_2, \nonumber\\
 && \wp(1|\hat{z},1)=\wp(1|\hat{z},3)=\wp(1|\hat{z},4)=0,\\
&& \tilde{\rho}^{\hat{x}}_0=\wp(0|\hat{x},3) \wp_3 \rho_3,\nonumber\\
&& \wp(0|\hat{x},1)=\wp(0|\hat{x},2)=\wp(0|\hat{x},4)=0,\\
&& \tilde{\rho}^{\hat{x}}_1=\wp(1|\hat{x},4) \wp_4 \rho_4,\nonumber\\
&& \wp(1|\hat{x},1)=\wp(1|\hat{x},2)=\wp(1|\hat{x},3)=0.
\end{eqnarray}

With the help of Eq. (\ref{const3}), one has
\begin{eqnarray}
\wp(0|\hat{z},1)=\wp(1|\hat{z},2)=\wp(0|\hat{x},3) =\wp(1|\hat{x},4) =1.
\end{eqnarray}
This directly yields
\begin{eqnarray}\label{E0-p4}
 \tilde{\rho}^{\hat{z}}_0&=&  \wp_1 \rho_1,\label{Erhoz0q-4}\nonumber\\
 \tilde{\rho}^{\hat{z}}_1&=& \wp_2 \rho_2,\label{Erhoz1q-4}\nonumber\\
 \tilde{\rho}^{\hat{x}}_0&=& \wp_3 \rho_3,\label{Erhox0q-4}\nonumber\\
 \tilde{\rho}^{\hat{x}}_1&=& \wp_4 \rho_4,\label{Erhox1q-4}
\end{eqnarray}
which is just the  set of equations given in (\ref{four}).
It can be verified that
\begin{eqnarray}
&& \tilde{\rho}^{\hat{z}}_0+
 \tilde{\rho}^{\hat{z}}_1=
 \tilde{\rho}^{\hat{x}}_0+
 \tilde{\rho}^{\hat{x}}_1=\rho_B,\nonumber\\
&& \rho_B={\rm
tr}_A(\rho_{AB})=\cos^2\theta |0\rangle\langle 0|+\sin^2\theta |1\rangle\langle 1|.
\end{eqnarray}
For Eq. (\ref{E0-p4}), by summing them up and taking trace,
the left-hand side gives $2 {\rm tr}\rho_B=2$. But the right-hand
side, by definition in Eq. (\ref{const1}), gives ${\rm tr}(\sum_{\xi=1}^4 \wp_{\xi} \rho_{\xi})={\rm tr}\rho_B=1$, this leads to the sharp contradiction ``2=1," as shown in the main text.

\vspace{3mm}

\noindent{\bf Existence of LHS model for the pure separable state.} Consider now, however, a pure separable state of two qubits
\begin{eqnarray}
 |\Psi\rangle=|0\rangle\otimes|\beta\rangle.\label{sep}
\end{eqnarray}
For this state, we shall show that a local hidden state model does exist. Without loss of generality, let Alice's two choices of projective measurements be
\begin{eqnarray}
\mathcal {P}^{\hat{n}_i}_a=|\varphi^i_a\rangle\langle\varphi^i_a|
\end{eqnarray}
with
\begin{eqnarray}
|\varphi^i_0\rangle&=&\cos\alpha_i|0\rangle+\sin\alpha_i|1\rangle,\nonumber\\
|\varphi^i_1\rangle&=&\sin\alpha_i|0\rangle-\cos\alpha_i|1\rangle.
\end{eqnarray}
By acting these projectors on the separable state (\ref{sep}), Bob's four conditional states are found to be
\begin{eqnarray}
\tilde{\rho}^{\hat{n}_1}_0&=&\cos^2\alpha_1|\beta\rangle\langle\beta|,\nonumber\\
\tilde{\rho}^{\hat{n}_1}_1&=&\sin^2\alpha_1|\beta\rangle\langle\beta|,\nonumber\\
\tilde{\rho}^{\hat{n}_2}_0&=&\cos^2\alpha_2|\beta\rangle\langle\beta|,\nonumber\\
\tilde{\rho}^{\hat{n}_2}_1&=&\sin^2\alpha_2|\beta\rangle\langle\beta|.
\end{eqnarray}
It then turns out that there exists a local hidden state model, with Alice's strategy based on a single hidden state, that could simulate
the above Bob's four conditional states:
\begin{eqnarray}
&\{\wp_1=1,\rho_1=|\beta\rangle\langle\beta|\},&\nonumber\\
&P(0|n_1, 1)=\cos^2\alpha_1,&\nonumber\\
&P(1|n_1, 1)=\sin^2\alpha_1,&\nonumber\\
&P(0|n_2, 1)=\cos^2\alpha_2,&\nonumber\\
&P(1|n_2, 1)=\sin^2\alpha_2.&
\end{eqnarray}
Thus, local hidden state model is possible for pure separable states.

{\bf Acknowledgements}

J.L.C. is supported by the National Basic Research Program (973
Program) of China under Grant No.\ 2012CB921900 and the National Natural Science Foundation of China
(Grant Nos.\ 11175089 and 11475089). H.Y.S is supported by Institute for
Information and Communications Technology Promotion
(IITP) grant funded by the Korea Government (MSIP)
(No. R0190-15-2028, Practical and Secure Quantum Key
Distribution). A.K.P is supported by the Special
Project of University of Ministry of Education of China and
the Project of K. P. Chair Professor of Zhejiang University of
China.

{\bf Author Contributions}

J.L.C. initiated the idea. J.L.C., H.Y.S., Z.P.X., and A.K.P. derived the results and wrote the manuscript. All authors reviewed the manuscript.

{\bf Additional Information}

\textbf{Competing financial interests:} The authors declare no
competing financial interests.


\end{document}